\documentclass{ws-procs975x65}
\usepackage{graphicx}
\usepackage{natbib}
\usepackage[flushleft]{threeparttable}
\usepackage{wrapfig}
\usepackage{color}
\def\beq{\begin{equation}}
\def\eeq{\end{equation}}

\begin{document}

\title{Will a nuclear stellar disk form in the galaxy Henize 2-10?}
\author{Manuel Arca-Sedda$^*$, and Roberto Capuzzo-Dolcetta$^{**}$}
\address{Dept. of Physics, University of Rome ``Sapienza'',\\
P.le A. Moro 5, Rome, I-00185, Italy\\
$^*$e-mail: manuel.arcasedda@uniroma1.it\\
$^{**}$e-mail: roberto.capuzzodolcetta@uniroma1.it}

\begin{abstract}
We present results of a set of $N$-body simulations to model the future evolution of the 11 young massive clusters hosted in the central region of the dwarf starburst galaxy Henize 2-10, which contains at its center a massive black hole with a mass $M_{\rm BH} \simeq 2\times 10^6$ M$_\odot$. Nuclear star clusters are present in a great quantity of galaxies of mass similar to Henize 2-10. Our results \citep{ASCD15} show that the orbital decay and merging of the Henize 2-10 clusters will likely lead to the formation of a nuclear star cluster with mass $M_{\rm NSC} \simeq 4-6 \times 10^6$ M$_\odot$ and effective radius $r_{\rm NSC} \simeq 4.1$ pc. Additionally, we found that this mechanism can lead to the formation of disky structures with global properties similar to those of nuclear stellar disks, which reside in many ``middle-weight'' galaxies. This work confirms and enlarge recent results \citep{ASCDS15} that indicate how nuclear star clusters  and super massive black holes are only partially  correlated, since the formation process of nuclear star clusters is poorly affected by a black hole of the size of that in Henize 2-10.
A new result is that nuclear star clusters and nuclear stellar disks may share the same formation path.
\end{abstract}


\bodymatter

\section{Introduction}

Most of galaxies with masses above $\sim 10^8$ M$_\odot$ host at their centre compact massive objects (CMOs), whose nature and mass depend on the properties of the host galaxy. In particular, galaxies with masses above $10^{10}$ M$_\odot$ contain, usually, super massive black holes (SMBHs) and/or nuclear stellar disks (NSDs), while lighter galaxies host nuclear star clusters (NSCs) \citep{scot}. Moreover, in many galaxies with masses in the range $10^{10}-10^{11}$ M$_\odot$ SMBHs and NSCs (or NSDs) may co-exist in the galactic centre. 
The existence of scaling relations connecting SMBHs or NSCs to their host properties might help in shedding light on the mechanisms of origin of these objects. For instance, Ferrarese et al. (2006) \cite{frrs} found that the relation between the NSC mass and the velocity dispersion of its host is similar to that for SMBHs. However, more recent studies based on a larger amount of NSC data have pointed out that such relation is shallower for NSCs than for SMBHs \citep{LGH,ASCD14b}. 
 These results seem to indicate that SMBHs and NSCs represent two different manifestations of CMOs whose particular nature depends on the galactic environment,
while their interaction and common evolution is still matter of debate. 
One possible formation mechanism for NSCs is the ``dry-merger'' scenario, in which star clusters orbitally decay toward the galactic centre under the action of dynamical friction (DF) \citep{TrOsSp,Dolc93,ASCD14}. Their subsequent merging lead to the formation of a dense stellar nucleus with characteristics comparable to those of observed galactic nuclei. In this scenario, the presence of a SMBH in the galactic centre may turn on tidal forces strong enough to disrupt the infalling clusters, thereby quenching the formation of a nucleus. On the other hand, theoretical and numerical modelling of the inspiral of star clusters have clearly shown that the environmental tidal forces are significant only when the central SMBH mass exceeds $10^8$ M$_\odot$ \citep{antonini13,ASCD14b,ASCDS15}.
In this framework, galaxies hosting several star clusters near their photometric centres represent an ideal laboratory to test the dry-merger scenario. One example is the Henize 2-10 starburst dwarf galaxy \citep{kobul95}. Indeed, this galaxy hosts an SMBH candidate of mass ${\rm Log} (M/{\rm M}_\odot)=6.3\pm 1.1$ \citep{merloni,reines}, and 11 young super-star clusters (SSCs) with masses above $10^5$ M$_\odot$, orbiting at projected distances $\lesssim 140$ pc from the galactic centre \citep{ngu14}. Detailed information available on the projected mass profile of the galaxy and on the SSC masses and sizes make Henize 2-10 an interesting case study for testing the dry-merger scenario. In particular, Arca-Sedda et al. (2015) \cite{ASCD15} used several direct $N$-body simulations to investigate whether the future evolution of the star cluster system (SCS) observed in Henize 2-10 will lead to the formation of a central stellar nucleus. The set of initial conditions chosen for the SSCs are consistent with the high-quality data available for their orbits and internal properties.

\section{Modelling Henize 2-10 and its star cluster system}

We shortly resume here the methods and results of the Arca-Sedda et al. (2015) \citep{ASCD15} paper. The galaxy density profile was modelled as:
\begin{equation}
\rho_\gamma(r)=\frac{(3-\gamma)M_H}{4\pi r_s^3}\left(\frac{r}{r_s}\right)^{-\gamma}\left(\frac{r}{r_s}+1\right)^{\gamma-4}\frac{1}{{\mathrm{cosh}(r/r_{\rm tr})}},
\end{equation}
where $M_H$ is the total mass of the galaxy, $r_s$ its scale radius, $\gamma$ is the inner density profile slope,
${\mathrm{cosh}(r/r_{\rm tr})}$ is the usual hyperbolic cosine function, and $r_{\rm tr}=150$ pc  is the truncation radius of the model. 
At radii $r \ll r_{\rm tr}$ the model converges to a Dehnen's density profile \citep{Deh93}.

This density profile represents satisfactorily the inner $150$ pc of Henize 2-10 under the choice of parameters: $M_H = 1.6\times 10^{9}$ M$_\odot$, $r_s = 110$ pc and $\gamma=1/2$. In particular, our model has an effective radius $R_e = 269$ pc and a velocity dispersion $\sigma = 41$ km s$^{-1}$ averaged in the region between $40$ and $260$ pc from the center, values very close to those inferred from observations. To highlight the role played by the SMBH in the formation of the NSC, we compared the results in 
models containing an SMBH with those obtained without an SMBH in the galaxy center.

To represent the eleven SSCs (hereafter labelled with the letter C followed by a number from 1 to 11) we adopted King models tuned on data provided by Nguyen et al. (2014) \cite{ngu14}. Furthermore, we chose three different sets of ICs for the SSCs motion, referred to as S1, S2 and S3 in the following. 
In configuration S1, the SSCs ICs are sampled from the same distribution function of the background galaxy, with the constraint that the clusters projected positions correspond to their actually observed positions. In S2, instead, all the SSCs move on circular orbits on the same plane. This scenario is motivated by recent observations claiming that the clusters move on the same disk of the molecular gas that resides in the Henize 2-10 nucleus. In this case, projected and spatial positions of the SSCs coincide and dynamical friction acts more efficiently. Finally, SSCs in configuration S3 have the same ICs as in S1, but in this case the galaxy does not contain a central SMBH. Finally, we present here also results coming from the choice of another configuration, S4, in which the SSCs lie initially on the same plane, but their velocity vectors do not.  

\section{Results from $N$-body modelling of the Henize 2-10 nucleus}

As stated above, we performed three simulations of the orbital evolution of the 11 SSCs traversing the inner $150$ pc of Henize 2-10. To do this, we used the direct $N$-body code HiGPUs \citep{Spera}. This code allows us to use more than $10^6$ ``particles'' to model the whole system (galaxy+SSCs) with a resolution sufficient to follow correctly the global internal dynamics of the SSCs.
The runs required $\sim 3$ months, producing over $1$ Tb of data, that have been deeply analysed using a well suited analysis tool that we developed in order to get all the relevant orbital and structural parameters of the SSCs and of the SMBH (if present) as a function of time. 
In the following, we will resume the main results obtained through this series of simulations. 

\subsection{Time for NSC assembly}
With semi-analytical estimates \citep{ASCD14b,ASCD14} we see that clusters C1-C4 (the heaviest of the 11) will reach the inner region of the galaxy in a short time ($20-80$ Myr), leading to a rapid accumulation of mass therein. The mass deposited within $20$ pc from the MBH is $M_{\rm dep}=4\times 10^6$ M$_\odot$ in configuration S1, $M_{\rm dep}=5\times 10^6$ M$_\odot$ in configuration S2, and  $M_{\rm dep}=4.5\times 10^6$ M$_\odot$ in configuration S3. The small difference between $M_{\rm dep}$ in models S1 and S3 ($11\%$) suggests that the presence of a central MBH   affects poorly the NSC build-up. As expected, $M_{\rm dep}$ is greater in model S2, because in this case all the SSCs reach the galactic centre. 
Actually, in the S2 case the decay process is quite fast, leading to the orbital decay of all the SSCs in $\sim 20$ Myr.

\subsection{Tidal erosion and dynamical friction}
As the SSCs move and decay orbitally within the galaxy, they undergo also tidal erosion (TE) operated by the galactic background and by the central MBH. This mechanism  may cause the erosion and disruption of the infalling clusters before they are able to reach the nucleus of the galaxy, thus suppressing the formation of a NSC. An analysis of our numerical results allowed us to discriminate among DF, TE arising from the galactic bakground, and TE ascribable to the central MBH. 
In all the cases considered, we found that the formation of a bright nucleus is essentially due to the decay of the four heaviest SSCs (C1-C4), with masses above $9\times 10^5$ M$_\odot$, while lighter (C5-C11) clusters are more affected by TE.
In model S1, the tidal forces exerted by the MBH enhance the SSC mass loss, such that they reach the centre of the galaxy with masses $\sim 15\%$ smaller than in model S3, in which the galaxy does not contain an MBH.
Furthermore, lighter SSCs are almost completely disrupted before they reach the galactic centre in both simulations S1 and S3. In these cases TE is mainly due to the galactic background, which halts the SSCs decay.
In model S2, DF acts more efficiently because SSCs have smaller apocentres than in the other two cases. In this case, all the SSCs reach the galactic centre and contribute significantly to the formation of a nucleus. 

\begin{table}
\tbl{NSC properties}
{\begin{tabular}{cccccccccc}
\toprule 
 & & & $M_{\rm NSC}$ & & & $r_{\rm NSC}$ & & & $r_{\rm eff}$ \\ 
  \colrule
S1 & & & $4.6$ & & & $10$ & & & $4.2$ \\  
S2 & & & $6.0$ & & & $10$ & & & $2.6$ \\  
S3 & & & $5.1$ & & & $ 6$ & & & $2.0$ \\  
\botrule 
\end{tabular}}
\label{T1}
\begin{tabnote}
Col. 1: model name. Col. 2: NSC mass in $10^6$ M$_\odot$. Col. 3: NSC radius in pc. Col. 4: NSC effective radius in pc.
\end{tabnote}
\end{table}

\subsection{Detectability of the newly born NSC}

Generally, a NSC is revealed as an edge in the surface luminosity profile of the host galaxy. As an example, in Fig. 1 we show the surface density profiles we obtain in case S2. In all the cases studied, we found a clear edge within 1 pc from the galaxy centre. In particular, we summarised in Table \ref{T1} the total mass, $M_{\rm NSC}$, radius, $r_{\rm NSC}$, and the effective radius, $r_{\rm eff}$, of the nucleated regions observed in our simulations.

\begin{figure}
\centering
\includegraphics[width=7cm]{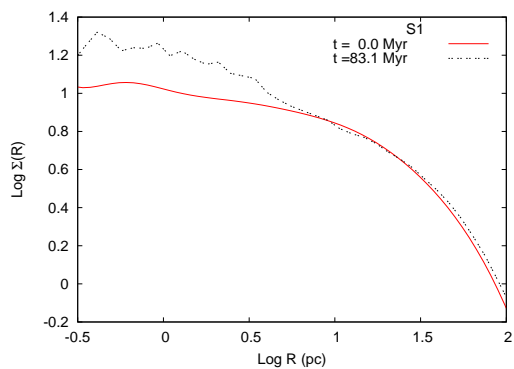}\\
\caption{Surface density profile of Henize 2-10 in configuration S2. The NSC is clearly visible within $\sim 1$ pc from the galactic centre.}
\label{F1}
\end{figure}

\subsection{When a NSC and when a NSD}
It is worth noting that nearly $20\%$ of the observed galaxies contains at their centres nuclear stellar disks (NSDs), quite independently on their Hubble type. NSDs have diameters of tens to hundreds pc, and masses in the range $10^7-10^9$ M$_\odot$.

Intriguingly, our results indicate that the clusters initial conditions are effective in determining the characteristics of the dense structure that forms in the galactic central region.

Actually, we found that the NSC has a nearly round shape in configurations S1 and S3, but it shows an evident disky structure in the case S2, in which all the SSC orbits lie on the same plane. In this case, we can speak of a NSD instead of a NSC.
To further investigate these points, we ran another simulation, S4, in which SSCs have initial positions as in model S2, but their initial velocity vectors are not co-planar. 
Even in this case, after the rapid orbital decay, a disky structure form, though its ``height-scale'' is twice that of the NSD grown in model S2.

\begin{figure}
\centering
\includegraphics[width=6cm]{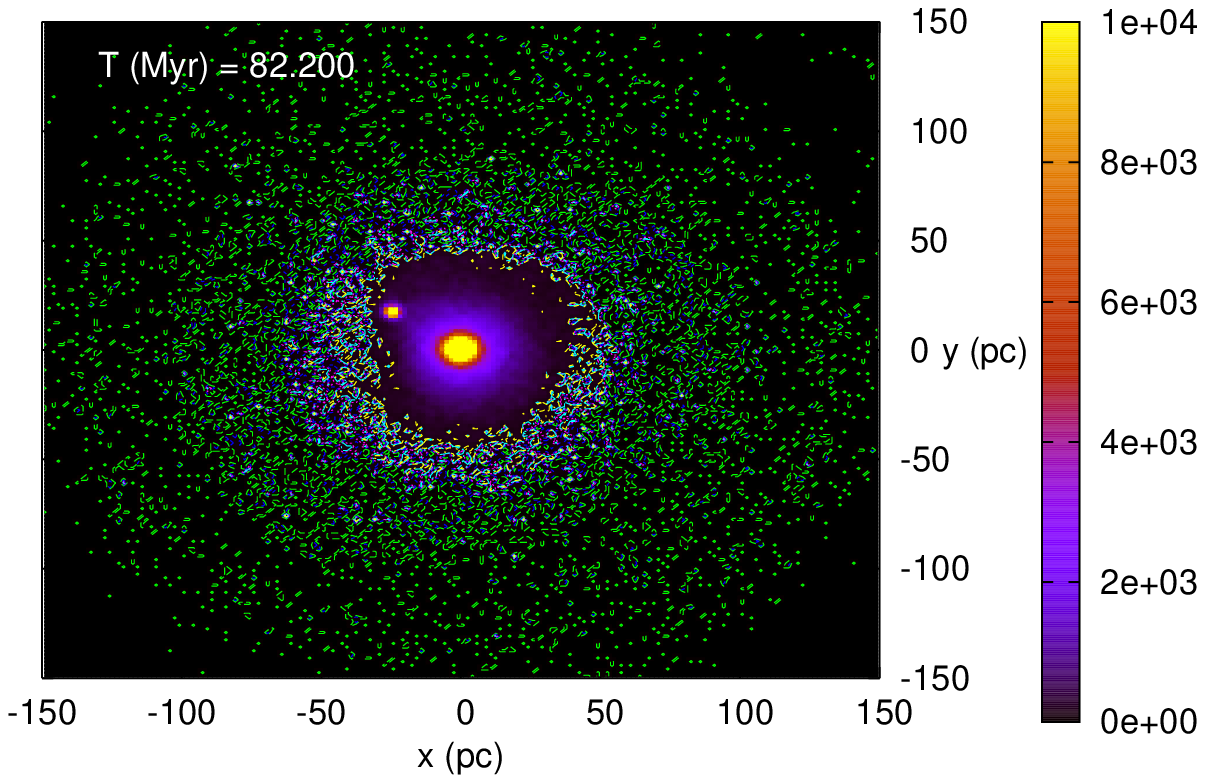}
\includegraphics[width=6cm]{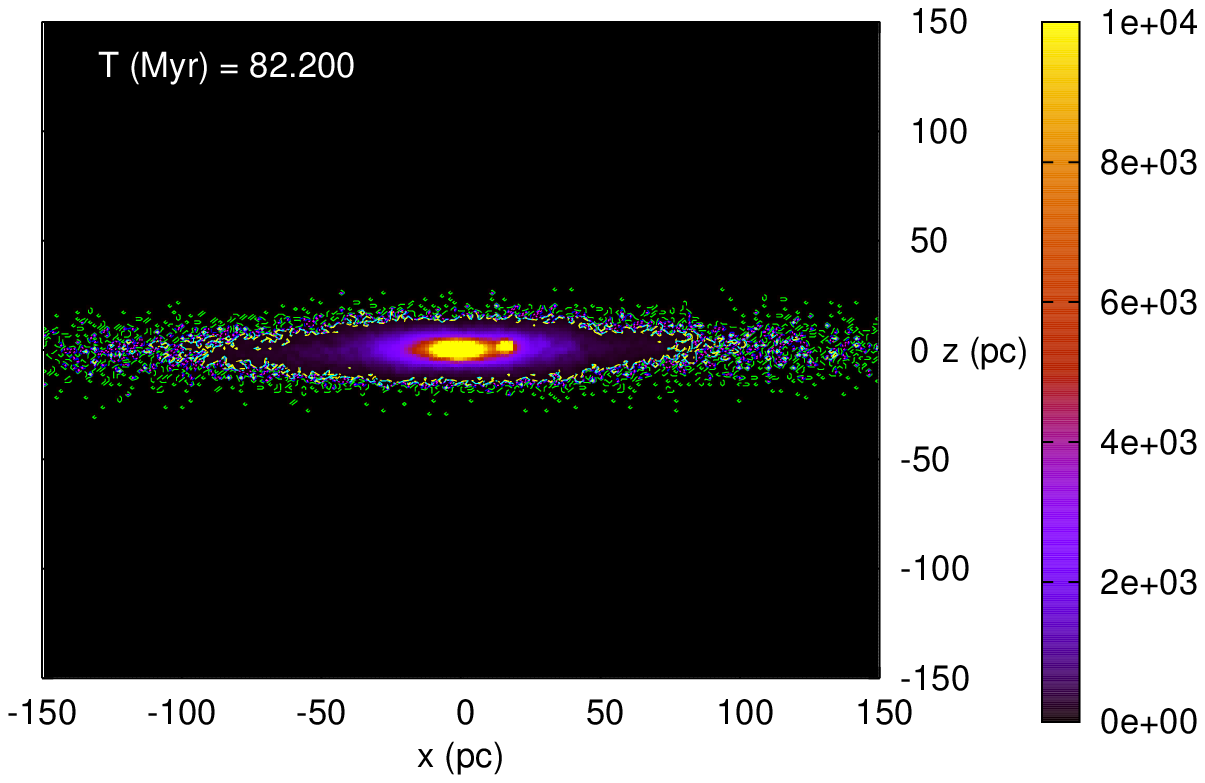}
\caption{Surface density maps of the SCS in model S2.}
\label{F2}
\end{figure}

Figure \ref{F2} shows the surface density maps of the SSCs in model S2 by the end of the simulation (after $\sim 80$ Myr) in the x-y and x-z plane. It is evident that the structure formed after the decay and merging of the 11 SSCs is a disk, with the symmetry axis in the z direction.

The disk is mainly composed of ``tidal debris'' lost by the SSCs during their orbital decay. The disk is characterised by an equatorial radius $R_{\rm d} = 100$ pc and a height-scale $z_{\rm d} \simeq 30$ pc.

Although some works proposed that NSDs originate in situ, our preliminary results suggest that the dry-merging scenario provides an alternative mode for their formation, which deserve to be further investigated.

\section{Conclusions}

In this work we have disucussed and enlarged previous results by Arca-Sedda et al. \citep{ASCD15} showing that the orbital decay of star clusters in a galaxy like the dwarf starbust Henize 2-10 can lead to the formation of a NSC although it hosts a pre-existent MBH. To perform this investigation, we used recent observational properties of the starburst galaxy Henize 2-10 as input for highly detailed, direct, $N$-body simulations. We found that not all the clusters are able to reach the centre, because the lightest of them are shattered by the tidal field of the parent galaxy, while the tidal effect caused by the central MBH slightly affects the quantity of mass deposited within the growing NSC. Therefore, it turns out that a MBH does not necessarily determine the properties of the NSC that surrounds it unless it is very massive. Furthermore, our results seem to suggest that the decay and merging of star clusters can represent also a satisfactory formation channel for nuclear stellar disk, though this depend strongly on their initial conditions.

%

\footnotesize{
\bibliographystyle{ws-procs975x65}
\bibliography{ASetal2015}
}

\end{document}